\documentclass[10pt,conference]{IEEEtran}
\usepackage{subfigure}
\usepackage{multirow}
\usepackage{graphicx}
\usepackage{amsmath}
\usepackage{color}
\usepackage{epsfig}
\usepackage{amsfonts}
\usepackage{amssymb}
\usepackage{amsthm}
\usepackage[usenames,dvipsnames]{pstricks}
\usepackage{epstopdf}
\usepackage{algorithm}
\usepackage[noend]{algpseudocode}

\newcommand{\argmax}{\mathop{\text{argmax}}}

\begin{document}

\title{{\LARGE MIMO-OTFS in High-Doppler Fading Channels: Signal Detection 
and Channel Estimation}}
\author{M. Kollengode Ramachandran and A. Chockalingam \\
Department of ECE, Indian Institute of Science, Bangalore 560012}
\maketitle

\begin{abstract}
Orthogonal time frequency space (OTFS) modulation is a recently introduced 
multiplexing technique designed in the 2-dimensional (2D) delay-Doppler 
domain suited for high-Doppler fading channels. OTFS converts a 
doubly-dispersive channel into an almost non-fading channel in the 
delay-Doppler domain through a series of 2D transformations. In this 
paper, we focus on MIMO-OTFS which brings in the high spectral and 
energy efficiency benefits of MIMO and the robustness of OTFS in 
high-Doppler fading channels. The OTFS channel-symbol coupling and 
the sparse delay-Doppler channel impulse response enable efficient 
MIMO channel estimation in high Doppler environments. We present an
iterative algorithm for signal detection based on message passing 
and a channel estimation scheme in the delay-Doppler domain suited for 
MIMO-OTFS. The proposed channel estimation scheme uses impulses in the 
delay-Doppler domain as pilots for estimation. We also compare the 
performance of MIMO-OTFS with that of MIMO-OFDM under high Doppler 
scenarios.  
\end{abstract}
\vspace{2mm}
{\em {\bfseries keywords:}}
{\em {\footnotesize OTFS modulation, MIMO-OTFS, 2D modulation, delay-Doppler 
domain, MIMO-OTFS signal detection, channel estimation.}} 

\section{Introduction}
\label{sec1}
Future wireless systems including 5G systems need to operate in dynamic 
channel conditions, where operation in high mobility scenarios (e.g., 
high-speed trains) and  millimeter wave (mm Wave) bands are envisioned. 
The wireless channels in such scenarios are doubly-dispersive, where 
multipath propagation effects cause time dispersion and Doppler shifts 
cause frequency dispersion \cite{jakes}. OFDM systems are usually  
employed to mitigate the effect of inter-symbol interference (ISI) 
caused by time dispersion \cite{ofdm1}. However, Doppler shifts result 
in inter-carrier interference (ICI) in OFDM and degrades performance 
\cite{ofdm2}. An approach to jointly combat ISI and ICI is to use pulse 
shaped OFDM systems \cite{pulse1}-\cite{pulse3}. Pulse shaped OFDM systems 
use general time-frequency lattices and optimized pulse shapes in the 
time-frequency domain. However, systems that employ the pulse shaping 
approach do not efficiently address the need to support high Doppler shifts.  

Orthogonal time frequency space (OTFS) modulation is a recently proposed 
multiplexing scheme \cite{otfswhitepaper}-\cite{otfs3} which meets the 
high-Doppler signaling need through a different approach, namely, 
{\em multiplexing the modulation symbols in the delay-Doppler domain} 
(instead of multiplexing symbols in time-frequency domain as in traditional 
modulation techniques such as OFDM). OTFS waveform has been shown to be
resilient to delay-Doppler shifts in the wireless channel. For example, 
OTFS has been shown to achieve significantly better error performance 
compared to OFDM for vehicle speeds ranging from 30 km/h to 500 km/h in 
4 GHz band, and that the robustness to high-Doppler channels (e.g.,
500 km/h vehicle speeds) is especially notable, as OFDM performance 
breaks down in such high-Doppler scenarios \cite{otfs2}. When OTFS 
waveform is viewed in the delay-Doppler domain, it corresponds to a 
2D localized pulse. Modulation symbols, such as QAM symbols, are 
multiplexed using these pulses as basis functions. The idea is to 
transform the time-varying multipath channel into a 2D time-invariant 
channel in the delay-Doppler domain. This results in a simple and 
symmetric coupling between the channel and the modulation symbols, due 
to which significant performance gains compared to other multiplexing 
techniques are achieved \cite{otfswhitepaper}. OTFS modulation can be 
architected over any multicarrier modulation by adding pre-processing and 
post-processing blocks. This is very attractive from an implementation 
view-point. 

Recognizing the promise of OTFS in future wireless systems, including
mmWave communication systems \cite{otfs3}, several works on OTFS have 
started emerging in the recent literature \cite{otfs4}-\cite{otfs9}. 
These works have addressed the formulation of input-output relation
in vectorized form, equalization and detection, and channel estimation. 
Multiple-input multiple-output (MIMO) techniques along with OTFS 
(MIMO-OTFS) can achieve increased spectral/energy efficiencies and 
robustness in rapidly varying MIMO channels. It is shown in 
\cite{otfswhitepaper} that OTFS approaches channel capacity through 
linear scaling of spectral efficiency with the MIMO order. We, in this 
paper, consider the signal detection and channel estimation aspects
in MIMO-OTFS.

\begin{figure*}[t]
\centering
\includegraphics[width=14.0 cm, height=4.0 cm]{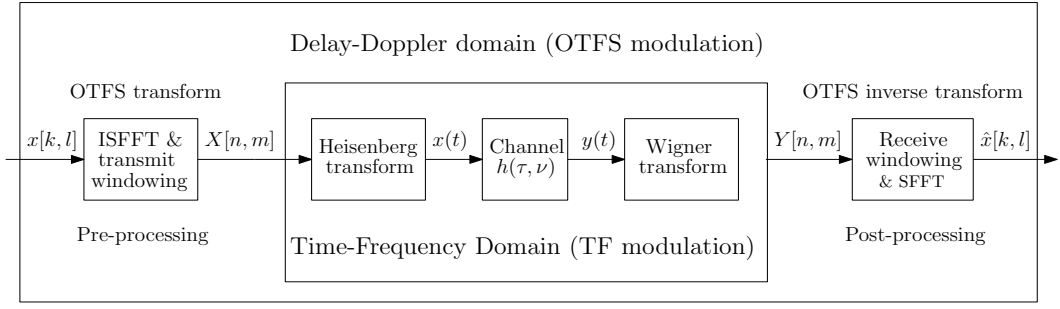}
\caption{OTFS modulation scheme.}
\label{fig2}
\vspace{-2mm}
\end{figure*}

Our contributions can be summarized as follows. We first present a 
vectorized input-output formulation for the MIMO-OTFS system. 
Initially, we assume perfect channel knowledge at the receiver and 
employ an iterative algorithm based on message passing for signal 
detection. The algorithm has low complexity and it achieves very good 
performance. For example, in a $2\times 2$ MIMO-OTFS system, a bit 
error rate (BER) of $10^{-5}$ is achieved at an SNR of about 14 dB 
for a Doppler of 1880 Hz (500 km/hr speed at 4 GHz). For the same 
system, MIMO-OFDM BER performance floors at a BER of 0.02. Next, we 
relax the perfect channel estimation assumption and present a channel 
estimation scheme in the delay-Doppler domain. The proposed scheme 
uses impulses in the delay-Doppler domain as pilots for MIMO-OTFS 
channel estimation. The proposed scheme is simple and effective in 
high-Doppler MIMO channels. For example, compared to the case of 
perfect channel knowledge, the proposed scheme loses performance 
only by less than a fraction of a dB.   

The rest of the paper is organized as follows. The OTFS modulation is
introduced in Sec. \ref{sec2}. The MIMO-OTFS system model and the 
vectorized input-output relation are developed in Sec. \ref{sec3}. 
MIMO-OTFS signal detection using message passing and the resulting BER
performance are presented in Sec. \ref{sec4}. The channel estimation 
scheme in the delay-Doppler domain and the achieved performance are 
presented in Sec. \ref{sec5}. Conclusions are presented in Sec. \ref{sec6}.

\section{OTFS Modulation}
\label{sec2}
OTFS modulation uses the delay-Doppler domain for multiplexing the 
modulation symbols and for channel representation. When the channel 
impulse response is represented in the delay-Doppler domain, the 
received signal $y(t)$ is the sum of reflected copies of the transmitted 
signal $x(t)$, which are delayed in time ($\tau$), shifted in frequency 
($\nu$), and multiplied by the complex gain $h(\tau,\nu)$ \cite{otfs1}. 
Thus, the coupling between an input signal and the channel in this domain 
is given by the following double integral:
\begin{equation}
\label{channel}
y(t)=\int_{\nu} \int_{\tau} h(\tau,\nu)x(t-\tau)e^{j2\pi\nu(t-\tau)} \mathrm{d} \tau \mathrm{d} \nu.
\end{equation}
The block diagram of the OTFS modulation scheme is shown in Fig. \ref{fig2}. 
The inner box is the familiar time-frequency multicarrier modulation, and 
the outer box with a pre- and post-processor implements the OTFS modulation 
scheme in the delay-Doppler domain. The information symbols $x[k,l]$ (e.g., 
QAM symbols) residing in the delay-Doppler domain are first transformed to 
the familiar time-frequency (TF) domain signal $X[n,m]$ through the 2D 
inverse symplectic finite Fourier transform (ISFFT) and windowing. The 
Heisenberg transform is then applied to the TF signal $X[n,m]$ to transform 
to the time domain signal $x(t)$ for transmission. At the receiver, the 
received signal $y(t)$ is transformed back to a TF domain signal $Y[n,m]$ 
through Wigner transform (inverse Heisenberg transform). $Y[n,m]$ thus 
obtained is transformed to the delay-Doppler domain signal $y[k,l]$ through 
the symplectic finite Fourier transform (SFFT) for demodulation.

In the following subsections, we describe the signal models in TF modulation 
and OTFS modulation. Let $T$ denote the TF modulation symbol time and 
$\Delta f$ denote the subcarrier spacing. Let $x[k,l]$, $k=0,\cdots,N-1$, 
$l=0,\cdots,M-1$ be the information symbols transmitted in a given packet 
burst. Let $W_{tx}[n,m]$ and $W_{rx}[n,m]$ denote the transmit and receive 
windows, respectively. 

\subsection{Time-frequency modulation }
\label{sec2a}
\begin{itemize}
\item Let  $\varphi_{tx}(t)$ and $\varphi_{rx}(t)$ denote the transmit and 
receive pulses, respectively, which are bi-orthogonal with respect to time 
and frequency translations. Signal in the TF domain $X[n,m]$, 
$n=0,\cdots,N-1$, $m=0,\cdots,M-1$ is transmitted in a given packet burst. 
\item TF modulation/Heisenberg transform: The signal in the time-frequency 
domain $X[n,m]$ is transformed to the time domain signal $x(t)$ using the 
Heisenberg transform given by
\begin{equation}
\hspace{-2mm}
x(t)= \sum_{n=0}^{N-1} \sum_{m=0}^{M-1} X[n,m]\varphi_{tx}(t-nT)e^{j2\pi m \Delta f (t-nT)}.
\label{tfmod}
\end{equation}
\item TF demodulation/Wigner transform: At the receiver, the time domain 
signal is transformed back to the TF domain using Wigner transform given by 
\begin{equation}
\label{wigner}
Y[n,m] = A_{\varphi_{rx},y}(\tau,\nu)|_{\tau =nT,\nu =m \Delta f},
\end{equation} 
where $A_{\varphi_{rx},y}(\tau,\nu)$ is the cross ambiguity function given by
\begin{equation}
\label{crossambig}
A_{\varphi_{rx},y}(\tau,\nu)=\int \varphi_{rx} ^*(t-\tau) y(t) e^{-j2 \pi \nu(t-\tau)} \mathrm{d}t,
\end{equation}
\end{itemize}
and $y(t)$ is related to $x(t)$ by (\ref{channel}). The relation between 
$Y[n,m]$ and $X[n,m]$ for TF modulation can be derived as \cite{otfs2}
\begin{equation}
\label{tfinpop}
Y[n,m] = H[n,m]X[n,m] + V[n,m],
\end{equation}
where $V[n,m]$ is the additive white Gaussian noise and $H[n,m]$ is given by
\begin{equation}
H[n,m]=\int_{\tau} \int_{\nu} h(\tau,\nu) e^{j2\pi \nu nT} e^{-j2\pi (\nu + m \Delta f) \tau} \mathrm{d} \nu \mathrm{d} \tau.
\end{equation}

\subsection{OTFS modulation}
\label{sec2b}
\begin{itemize}
\item 	Let $X_p[n,m]$ be the periodized  version of $X[n,m]$ with period 
	$(N,M)$. The SFFT of  $X_p[n,m]$ is given by
\begin{equation}
x_p[k,l] = \sum_{n=0}^{N-1} \sum_{m=0}^{M-1} X_p[n,m] e^{-j2\pi( {nk \over N} - {ml \over M} )},
\end{equation}
and the ISFFT is {\small $X_p[n,m] = SFFT^{-1} (x[k,l])$}, given by
\begin{equation}
X_p[n,m] = {1 \over MN }\sum_{k=0}^{N-1} \sum_{l=0}^{M-1} x[k,l] e^{j2\pi( {nk \over N}-{ml \over M})}.
\end{equation}
\item Information symbols $x[k,l]$, $k=0,\cdots,N-1$, $l=0,\cdots,M-1$, 
are transmitted in a given packet burst.
\item OTFS transform/pre-processing: The information symbols in the 
delay-Doppler domain $x[k,l]$ are mapped to TF domain symbols $X[n,m]$ as 
\begin{equation}
X[n,m] = W_{tx}[n,m]SFFT^{-1}(x[k,l]),
\label{otfsmod}
\end{equation}
where $W_{tx}[n,m]$ is the transmit windowing square summable function. 
\item $X[n,m]$ thus obtained is in the TF domain and it is TF modulated 
as described in the previous subsection, and $Y[n,m]$ is obtained by  
(\ref{wigner}).
\item OTFS demodulation/post-processing: A receive window $W_{rx}[n,m]$ is 
applied to $Y[n,m]$ and periodized to obtain $Y_p[n,m]$ which has the period  
$(N,M)$, as 
\begin{eqnarray}
Y_W[n,m] & = & W_{rx}[n,m]Y[n,m],  \nonumber \\
Y_p[n,m] & = & \sum_{k,l=- \infty}^{\infty}  Y_W[n-kN,m-lM].
\label{otfsdemod1}
\end{eqnarray}
The symplectic finite Fourier transform is then applied to $Y_p[n,m]$ to 
convert it from TF domain back to delay-Doppler domain $\hat{x}[k,l]$, as
\begin{equation}
\hat{x}[k,l]=SFFT (Y_p[n,m]).
\label{otfsdemod2}
\end{equation}
\end{itemize}
The input-output relation in OTFS modulation can be derived 
as \cite{otfs2}
\begin{equation}
\hat{x}[k,l]={1 \over MN} \hspace{-1mm} \sum_{m=0}^{M-1} \sum_{n=0}^{N-1}\hspace{-1.0mm} x[n,m] h_w \hspace{-1.0mm} \left( {k-n \over NT}, {l-m \over M \Delta f} \right)\hspace{-1mm} + v[k,l],
\label{otfsinpoutp}
\end{equation}
where
\begin{equation}
h_w \left({k-n \over NT}, {l-m \over M \Delta f} \right) = h_w (\nu',\tau')|_{\nu'={k-n \over NT},\tau'={l-m\over M \Delta f}},
\label{deldoppchannel}
\end{equation}
where
$h_w(\nu',\tau')$ is the circular convolution of the channel response 
with a windowing function $w(\tau,\nu)$, given by
\begin{equation}
h_w(\nu',\tau')=\int_{\nu} \int_{\tau} h(\tau,\nu)w(\nu'-\nu,\tau'-\tau) \mathrm{d} \tau \mathrm{d} \nu,
\end{equation}
where $w(\tau,\nu)$ is given by
\begin{equation}
w(\tau,\nu) = \hspace{-1mm} \sum_{m=0}^{M-1} \sum_{n=0}^{N-1} \hspace{-1mm} W_{tx}[n,m] W_{rx}[n,m] e^{-j2 \pi (\nu nT - \tau m \Delta f)}. 
\end{equation}

\begin{figure*}[t]
\centering
\includegraphics[width=16 cm, height=4.3cm]{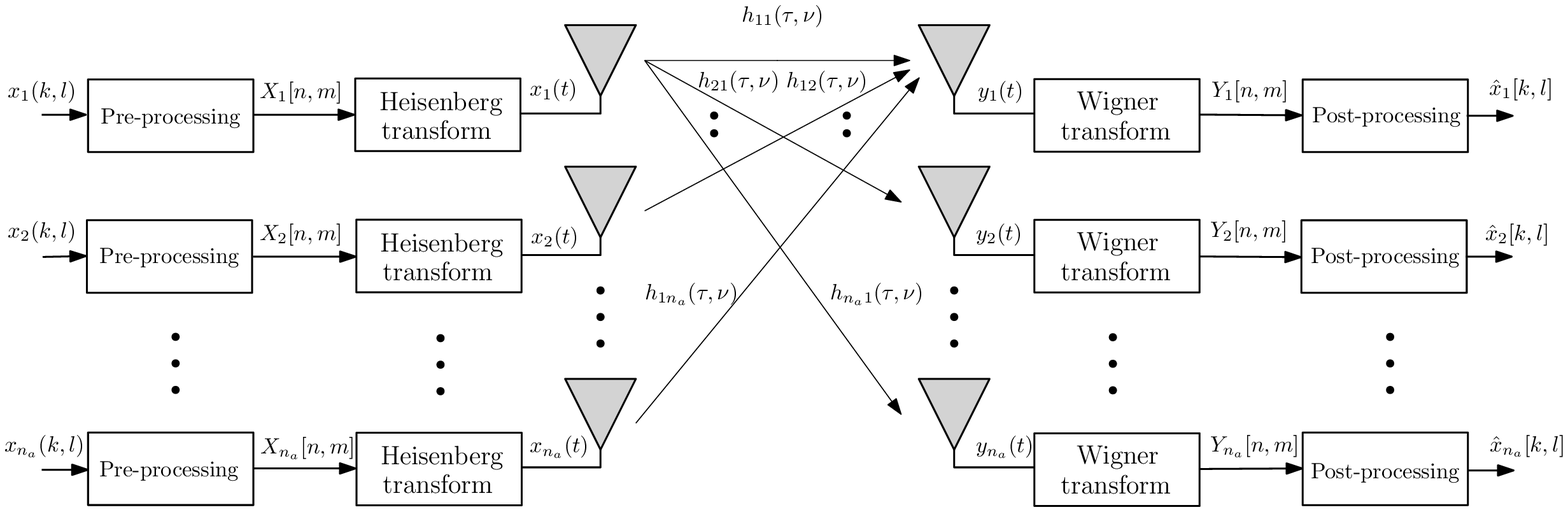}
\caption{MIMO-OTFS modulation scheme.}
\label{MIMO_BD}
\vspace{-4mm}
\end{figure*}

\vspace{-2mm}
\subsection{Vectorized formulation of the input-output relation}
\label{sec2c}
Consider a channel with $P$ signal propagation paths (taps). Let the path 
$i$ be associated with a delay $\tau_i$, a Doppler $\nu_i$, and a fade 
coefficient $h_i$. The channel impulse response in the delay-Doppler 
domain can be written as
\begin{equation}
h(\tau,\nu) =\sum_{i=1}^{P} h_i \delta(\tau -\tau_i) \delta(\nu-\nu_i).
\label{sparsechannel}
\end{equation}
Assume that the windows used in modulation, $W_{tx}[n,m]$ and $W_{rx}[n,m]$ 
are rectangular. Define $\tau_i={ \alpha _i \over M \Delta f}$ and 
$\nu_i={\beta_i \over NT}$, where $\alpha_i$ and $\beta_i$ are 
integers denoting the indices of the delay tap (with delay $\tau_i$) and 
Doppler tap (with Doppler value $\nu_i$). In practice, although the delay 
and Doppler values are not exactly integer multiples of the taps, they can 
be well approximated by a few delay-Doppler taps in the discrete domain 
\cite{channelestimation1}. With the above assumptions, the input-output 
relation for the channel in (\ref{sparsechannel}) can be derived as 
\cite{otfs5}
\begin{equation}
y[k,l] = \sum_{i=1}^{P} h_i' x[((k-\beta_i))_N,((l-\alpha_i))_M)] + v[k,l]. 
\label{inpopnofracdopp}
\end{equation}
where $h_i'=h_i e^{-j2 \pi \nu_i \tau_i}$. The above equation 
can be represented in vectorized form as \cite{otfs5}
\begin{equation}
\mathbf{y} = \mathbf{Hx} + \mathbf{v}, 
\label{vecform}
\end{equation}
where $\mathbf{x}, \mathbf{y}, \mathbf{v} \in \mathbb{C} ^{NM \times 1}$, 
$\mathbf{H} \in \mathbb{C}^{NM\times NM}$, the $(k + Nl)$th element of 
$\mathbf{x}$,  $x_{k+Nl}=x[k,l]$, $k=0,\cdots,N-1, l=0,\cdots,M-1$, and 
the same relation holds for $\mathbf{y}$ and $\mathbf{z}$ as well. In 
this representation, there are only $P$ non-zero elements in each row and 
column of the equivalent channel matrix ($\mathbf{H}$) due to modulo 
operations. 

\section{MIMO-OTFS Modulation}
\label{sec3}
Consider a MIMO-OTFS system as shown in Fig. \ref{MIMO_BD}  with equal 
number of transmit ($n_t$) and receive antennas ($n_r$), i.e., 
$n_t=n_r=n_a$. Each antenna transmits  OTFS modulated information symbols 
independently. Let the windows $W_{tx}[n,m]$, $W_{rx}[n,m]$ used for 
modulation be rectangular. Assume that the channel corresponding to 
$p$th transmit antenna and $q$th receive antenna has $P$ taps as in 
(\ref{sparsechannel}). Therefore, the channel representation can be 
written as  
\begin{equation}
h_{qp}(\tau,\nu)=\sum_{i=1}^{P} h_{qp_i} \delta(\tau-\tau_i) \delta(\nu-\nu_i),
\label{sparsechannelmimo}
\end{equation} 
$p=1,2,\cdots,n_a$, $q=1,2,\cdots,n_a$. Thus, we can use the vectorized 
formulation in Sec. \ref{sec2c} for each transmit and receive antenna 
pair to describe the input-output relation. 

\subsection{Vectorized formulation of the input-output relation for MIMO-OTFS}
\label{sec3a}
Let $\mathbf{H}_{qp}$ denote the equivalent channel matrix corresponding 
to $p$th transmit antenna and $q$th receive antenna. Let $\mathbf{x}_p$ 
denote the $NM \times 1$ transmit vector from the $p$th transmit antenna 
and $\mathbf{y}_q$ denote the $NM \times 1$ received vector corresponding 
to $q$th receive antenna in a given frame. Then, similar to the system 
model in (\ref{vecform}) for a SISO-OTFS, we can derive a linear system 
model describing the input and output for the MIMO-OTFS system as given below
\begin{eqnarray}
\mathbf{y}_1 & = & \mathbf{H}_{11}\mathbf{x}_1+\mathbf{H}_{12}\mathbf{x}_2 + \cdots + \mathbf{H}_{1{n_a}}\mathbf{x}_{n_a}+\mathbf{v}_1,   \nonumber \\
\mathbf{y}_2 & = & \mathbf{H}_{21}\mathbf{x}_1+\mathbf{H}_{22}\mathbf{x}_2 + \cdots + \mathbf{H}_{2{n_a}}\mathbf{x}_{n_a}+\mathbf{v}_2,\nonumber \\
\vdots \nonumber \\
\hspace{-4mm}\mathbf{y}_{n_a} & = & \mathbf{H}_{{n_a}1}\mathbf{x}_1+\mathbf{H}_{{n_a}2}\mathbf{x}_2 + \cdots + \mathbf{H}_{{n_a}{n_a}}\mathbf{x}_{n_a}+\mathbf{v}_{n_a}.\hspace{4mm}
\label{mimoeqns}
\end{eqnarray}
Define
\vspace{-3mm}
\begin{align*} 
\  \mathbf{H}_{ {\tiny \mbox{MIMO}}} &= \begin{bmatrix}
\mathbf{H} _{11} & \mathbf{H}_{12} & \dots & \mathbf{H}_{1{n_a}} \\
\mathbf{H} _{21} & \mathbf{H}_{22} & \dots & \mathbf{H}_{2{n_a}} \\
\vdots & \vdots & \ddots & \vdots \\
\mathbf{H} _{{n_a}1} & \mathbf{H}_{{n_a}2} & \dots & \mathbf{H}_{{n_a}{n_a}} 
\end{bmatrix}, 
\end{align*}

\vspace{-3mm}
\begin{small}
\begin{align*}
&\mathbf{x}_{{\tiny \mbox{MIMO}}}={[{\mathbf{x}_1}^{T},{\mathbf{x}_2}^{T},\cdots,  {\mathbf{x}_{n_a}}^{T}] }^{T}, \mathbf{y}_{{\tiny \mbox{MIMO}}} = {[{\mathbf{y}_1}^{T},{\mathbf{y}_2}^{T},\cdots,  {\mathbf{y}_{n_a}}^{T}] }^{T}, \\ & \mathbf{v}_{{\tiny \mbox{MIMO}}} = {[{\mathbf{v}_1}^{T},{\mathbf{v}_2}^{T},\cdots,  {\mathbf{v}_{n_a}}^{T}] }^{T}.
\end{align*}
\end{small}         
Then, (\ref{mimoeqns}) can be written as
\begin{equation}
\mathbf{y}_{{\tiny \mbox{MIMO}}} = \mathbf{H}_{{\tiny \mbox{MIMO}}}\mathbf{x}_{{\tiny \mbox{MIMO}}} + \mathbf{v}_{\tiny \mbox{MIMO}}, 
\label{mimovecform}
\end{equation}
where 
$\mathbf{x}_{{\tiny \mbox{MIMO}}}, \mathbf{y}_{{\tiny \mbox{MIMO}}}, \mathbf{v}_{\tiny \mbox{MIMO}} \in \mathbb{C} ^{{n_a}NM \times 1}$, 
$\mathbf{H}_{{\tiny \mbox{MIMO}}} \in \mathbb{C}^{{n_a}NM\times {n_a}NM}$. 
Thus, in this representation, each row and column of 
$\mathbf{H}_{{\tiny \mbox{MIMO}}}$ has only $n_aP$ non-zero elements 
due to modulo operations.
  
\section{MIMO-OTFS Signal Detection}
\label{sec4}
In this section, we present a MIMO-OTFS signal detection scheme using an 
iterative algorithm based on message passing and present a performance 
comparison between MIMO-OTFS and MIMO-OFDM in high-Doppler scenarios.

\subsection{Algorithm for MIMO-OTFS signal detection}
\label{sec4a}
Let the sets of non-zero positions in the $b$th row and $a$th column of 
$\mathbf{H}_{{\tiny \mbox{MIMO}}}$ be denoted by $\zeta_b$ and $\zeta_a$, 
respectively. Using (\ref{mimovecform}), the system can be modeled as a 
sparsely connected factor graph with ${n_a}NM$ variable nodes corresponding 
to the elements in $\mathbf{x}_{{\tiny \mbox{MIMO}}}$ and ${n_a}NM$ 
observation nodes corresponding to  the elements in 
$\mathbf{y}_{{\tiny \mbox{MIMO}}}$. Each observation node $y_b$ is 
connected to the set of variable nodes \{$x_c, c \in \zeta_b$\}, and each 
variable node $x_a$ is connected to the set of observation nodes 
\{$y_c, c \in \zeta_a$\}. Also, $|\zeta_b|=|\zeta_a|={n_a}P$. The maximum 
a posteriori (MAP) decision rule for (\ref{mimovecform}) is given by
\begin{equation}
{\hat{\mathbf{x}}_{{\tiny \mbox{MIMO}}}} =  \argmax_{\mathbf{x}_{{\tiny \mbox{MIMO}}} \in \mathbb{A} ^ {n_aNM}} \mbox{Pr}(\mathbf{x}_{{\tiny \mbox{MIMO}}}|\mathbf{y}_{{\tiny \mbox{MIMO}}},\mathbf{H}_{{\tiny \mbox{MIMO}}}),
\label{mapx}
\end{equation}
where $\mathbb{A}$ is the modulation alphabet (e.g., QAM) used. The 
detection as per (\ref{mapx}) has exponential complexity. Hence, we use 
symbol by symbol MAP rule for $ 0 \leq a \leq n_aNM-1 $ for detection 
as follows: 
\begin{equation}
\label{MAPsbsmimo}
\begin{split}
\hat{x}_a & = \argmax_{a_j \in \mathbb{A}} \mbox{Pr}(x_a = a_j | \mathbf{y}_{{\tiny \mbox{MIMO}}},\mathbf{H}_{{\tiny \mbox{MIMO}}}) \\
          & = \argmax_{a_j \in \mathbb{A}} {1 \over |\mathbb{A}|} \mbox{Pr}(\mathbf{y}_{{\tiny \mbox{MIMO}}}|x_a =a_j , \mathbf{H}_{{\tiny \mbox{MIMO}}}  ) \nonumber \\
          & \approx \argmax_{a_j \in \mathbb{A}} \prod_{c \in \zeta_a} \mbox{Pr}(y_c|x_a =a_j , \mathbf{H}_{{\tiny \mbox{MIMO}}}).
\end{split}
\end{equation}
The transmitted symbols are assumed to be equally likely and the components 
f $\mathbf{y}_{{\tiny \mbox{MIMO}}}$ are nearly independent for a given 
$x_a$ due to the sparsity in $\mathbf{H}_{{\tiny \mbox{MIMO}}}$. This can 
be solved using the message passing based algorithm described below. The 
message that is passed from the variable node $x_a$, for each 
$a= \{0,1, \cdots,n_aNM-1\}$, to the observation node $y_b$ for 
$b \in \zeta_a$, is the pmf denoted by 
$\textbf{p}_{ab} = \{p_{ab}(a_j)|a_j \in \mathbb{A} \}$ of the symbols in 
the constellation $\mathbb{A}$. Let $H_{ab}$ denote the element in the 
$a$th row and $b$th column of $\mathbf{H}_{{\tiny \mbox{MIMO}}}$. The 
message passing algorithm is described as follows.

\begin{algorithmic}[1]
\label{alg1}
\State \textbf{Inputs}: $\mathbf{y}_{{\tiny \mbox{MIMO}}}$, $\mathbf{H}_{{\tiny \mbox{MIMO}}}$,  $ N_{iter}$: max. number of iterations.
\State \textbf{Initialization}: Iteration index $t=0$, pmf $\mathbf{p}_{ab}^{(0)}=1/|\mathbb{A}| \  \forall \  a \in  \{0,1,\cdots,n_aNM-1 \}$ and $b \in \zeta_a $.
\State \textbf{Messages from $y_b$ to $x_a$}:  The mean $(\mu_{ba}^{(t)})$ 
and variance $((\sigma_{ba}^{(t)})^2)$ of the interference term $I_{ba}$ 
are passed as messages from $y_b$ to $x_a$.  $I_{ba}$ can be approximated 
as a Gaussian random variable and is given by

\begin{small}
\begin{equation}
I_{ba}=  \sum_{c \in \zeta_b,c \neq a } x_cH_{b,c} + v_b .
\end{equation}
\end{small}
The mean and variance of $I_{ba}$ are given by
\begin{small}
\begin{equation*}
\mu_{ba}^{(t)} = \mathbb{E}[I_{ba}] = \sum_{c \in \zeta_b, c \neq a } \sum_{j=1}^{|\mathbb{A}|} p_{cb}^{(t)}(a_j) a_j H_{b,c},
\end{equation*}
\begin{align*}
&( \sigma_{ba}^{(t)})^2 = \text{Var}[I_{ba}] \nonumber \\ 
&= \sum_{\substack{c \in \zeta_b \\ c \neq a}} \Bigg( \sum_{j=1}^{\mathbb{|A|}}p_{cb}^{(t)}(a_j)|a_j|^2 |H_{b,c}|^2 - \bigg| \sum_{j=1}^{\mathbb{|A|}}p_{cb}^{(t)}(a_j)a_jH_{b,c} \bigg|^2 \Bigg)
\end{align*}
\hspace{4.5mm} $+ \ \sigma^2.$
\end{small}

\State \textbf{Messages from $x_a$ to $y_b$}: Messages passed from variable 
nodes $x_a$ to observation nodes $y_b$ is the pmf vector 
$\textbf{p}_{ab}^{(t+1)}$ with the elements given by
\begin{equation}
p_{ab}^{(t+1)}=\Delta \  p_{ab}^{(t)}(a_j)+(1- \Delta) \  p_{ab}^{(t-1)}(a_j),
\end{equation}
where $\Delta \in (0,1]$ is the damping factor for improving convergence 
rate, and
\begin{equation}
p_{ab}^{(t)} \propto \prod_{c \in \zeta_a , c \neq b} \text{Pr}(y_c|x_a=a_j,\mathbf{H}_{{\tiny \mbox{MIMO}}}),
\end{equation}
where 
\begin{equation*}
\text{Pr}(y_c|x_a=a_j,\mathbf{H}_{{\tiny \mbox{MIMO}}}) \propto \text{exp} \Bigg( {-|y_c - \mu_{ca}^{(t)}-H_{c,a}a_j|^2 \over \sigma_{c,a}^{2(t)}}  \Bigg).
\end{equation*}

\State \textbf{Stopping criterion}: Repeat steps 3 \& 4 till 
$\max\limits_{a,b,a_j} |p_{ab}^{(t+1)}(a_j) - p_{ab}^{(t)}(a_j)| < \epsilon$ 
(where $\epsilon$ is a small value) or the maximum number of iterations, 
$N_{iter}$, is reached.
\State \textbf{Output}: Output the detected symbol as
\begin{equation}
\hat{x}_a = \argmax_{a_j \in \mathbb{A}}p_a(a_j), \:\: a \in {0,1,2,\cdots,n_aNM-1},
\end{equation}
where
\begin{equation}
p_a(a_j) = \prod_{c \in \zeta_a} \text{Pr}(y_c|x_a =a _j,\mathbf{H}_{{\tiny \mbox{MIMO}}}).
\end{equation}
\end{algorithmic}

\subsection{Vectorized formulation of the input-output relation for MIMO-OFDM}
\label{subsec4b}
In this subsection, in order to provide a performance comparison between 
MIMO-OTFS and MIMO-OFDM, we present the vectorized formulation of the 
input-output relation for MIMO-OFDM. OFDM uses the TF domain for signaling 
and channel representation. We will first derive the vectorized formulation 
for a SISO-OFDM and extend it to MIMO-OFDM. For a fair comparison with the 
OTFS modulation, we will consider $N$ consecutive OFDM blocks (each of size 
$M$) to be one frame, i.e., the transmit vector 
$\mathbf{x}_{{\tiny \mbox{OFDM}}} \in \mathbb{C} ^{NM \times 1}$, and 
message passing detection is done jointly over one $NM \times 1$ frame. 
Consider the channel in (\ref{sparsechannel}). The time-delay 
representation $h(\tau,t)$ is related to the delay-Doppler representation 
$h(\tau,\nu)$ by a Fourier transform along the time axis, and is given by 
\begin{equation}
h(\tau,t) =\sum_{i=1}^{P} h_i e^{j2 \pi \nu_i t}\delta(\tau-\tau_i).
\label{tdrep}
\end{equation}
Sample the time axis at $t = nTs ={n \over M\Delta f}$. The sampled 
time-delay representation $h(\tau,n)$ is given by
\begin{equation}
h(\tau,n) =\sum_{i=1}^{P} h_i e^{j2 \pi \nu_i n \over  M\Delta f}\delta(\tau-\tau_i).
\label{tdrep_discrete}
\end{equation}
Let $CP=P-1$ denote the cyclic prefix length used in each OFDM block and 
let $L=M+CP$. The size of one frame after cyclic prefix insertion to each 
block will then be $NL$. Let 
$\mathbf{T}_{CP}={[ \mathbf{C}_{CP}^T \ \mathbf{I}_M ]}^T$ denote the 
$L \times M$  matrix that inserts cyclic prefix for one block, where 
$\mathbf{C}_{CP}$ contains the last $CP$ rows of the identity matrix 
$\mathbf{I}_M$. Also, let 
$\mathbf{R}_{CP}= [ {\bf 0}_{M \times CP } \ {\mathbf I}_M ]$ denote the 
$M \times L$ the matrix that removes the cyclic prefix for one block 
\cite{rapid_tv_channels}. Let $\mathbf{W}_{M \times M}$ and 
$\mathbf{W}^{H}_{M \times M}$ denote the DFT and IDFT matrices of size $M$.
We use the following notations.
\begin{itemize}
\item $\mathbf{B}_{cpin}= \text{diag} \underbrace{(\mathbf{T}_{CP},\mathbf{T}_{CP},\cdots,\mathbf{T}_{CP})}_{N \ times}$ 
: cyclic prefix insertion matrix for $N$ consecutive OFDM blocks.
\item $\mathbf{B}_{cpre}= \text{diag} \underbrace{(\mathbf{R}_{CP},\mathbf{R}_{CP},\cdots,\mathbf{R}_{CP})}_{N \ times}$ 
: cyclic prefix removal matrix for $N$ consecutive OFDM blocks. 
\item $\mathbf{D}= \text{diag} \underbrace{(\mathbf{W},\mathbf{W},\cdots,\mathbf{W})}_{N \ times}$ 
: DFT matrix for $N$ consecutive OFDM blocks.
\item $\mathbf{D}^{H}= \text{diag} \underbrace{(\mathbf{W}^{H},\mathbf{W}^{H},\cdots,\mathbf{W}^{H})}_{N \ times}$ 
: IDFT matrix for $N$ consecutive OFDM blocks.
\item The channel in the time-delay domain for a given frame can be written 
as a matrix $\mathbf{H}_{td}$ using (\ref{tdrep_discrete}) and has size 
$NL \times NL$.
\end{itemize}
Using the above, the end-to-end relationship in OFDM modulation can be 
described by the following linear model:
\begin{align}
\mathbf{y}_{{\tiny \mbox{OFDM}}} & = \underbrace{\mathbf{D} \mathbf{B}_{cpre} \mathbf{H}_{td} \mathbf{B}_{cpin} \mathbf{D}^H}_{\mathbf{H}_{{\tiny \mbox{OFDM}}}} \mathbf{x}_{{\tiny \mbox{OFDM}}} + \mathbf{v} \nonumber \\ 
&= \mathbf{H}_{{\tiny \mbox{OFDM}}}  \mathbf{x}_{{\tiny \mbox{OFDM}}} + \mathbf{v},
\end{align}
where  
$\mathbf{x}_{{\tiny \mbox{OFDM}}}, \mathbf{y}_{{\tiny \mbox{OFDM}}}, \mathbf{v} \in \mathbb{C} ^{NM \times 1}$, 
$\mathbf{H}_{{\tiny \mbox{OFDM}}} \in \mathbb{C}^{NM\times NM}$.

\subsubsection{MIMO-OFDM}
The vectorized formulation of the input-output relation for SISO-OFDM 
derived above can be extended to MIMO-OFDM in a similar fashion as was done 
for the MIMO-OTFS system described in Sec. \ref{sec3a}. Let 
$\mathbf{H}_{{{\tiny \mbox{OFDM}}}_{qp}}$ denote the equivalent channel 
matrix corresponding to $p$th transmit antenna and $q$th receive antenna. 
Let $\mathbf{x}_{{{\tiny \mbox{OFDM}}}_p}$ denote the $NM \times 1$ transmit 
vector from the $p$th transmit antenna and 
$\mathbf{y}_{{{\tiny \mbox{OFDM}}}_q}$ denote the $NM \times 1$ received 
vector corresponding to $q$th receive antenna in a given frame. Define
\begin{align*} 
\  \mathbf{H}_{ {\tiny \mbox{MIMO-OFDM}}} &= \begin{bmatrix}
\mathbf{H} _{{\tiny \mbox{OFDM}}_{11}} & \mathbf{H}_{{\tiny \mbox{OFDM}}_{12}} & \dots & \mathbf{H}_{{\tiny \mbox{OFDM}}_{1{n_a}}} \\
\mathbf{H} _{{\tiny \mbox{OFDM}}_{21}} & \mathbf{H}_{{\tiny \mbox{OFDM}}_{22}} & \dots & \mathbf{H}_{{\tiny \mbox{OFDM}}_{2n_a}} \\
\vdots & \vdots & \ddots & \vdots \\
\mathbf{H} _{{\tiny \mbox{OFDM}}_{n_a1}} & \mathbf{H}_{{\tiny \mbox{OFDM}}_{n_a2}} & \dots & \mathbf{H}_{{\tiny \mbox{OFDM}}_{n_a n_a}} 
\end{bmatrix}, 
\end{align*}

\begin{small}
\begin{align*}
&\mathbf{x}_{{\tiny \mbox{MIMO-OFDM}}}={[{\mathbf{x}_{{\tiny \mbox{OFDM}}_1}}^{T},{\mathbf{x}_{{\tiny \mbox{OFDM}}_2}}^{T},\cdots,{\mathbf{x}_{{\tiny \mbox{OFDM}}_{n_a}}}^{T}]}^{T}, \\ & \mathbf{y}_{{\tiny \mbox{MIMO-OFDM}}} = {[{\mathbf{y}_{{\tiny \mbox{OFDM}}_1}}^{T},{\mathbf{y}_{{\tiny \mbox{OFDM}}_2}}^{T},\cdots,{\mathbf{y}_{{\tiny \mbox{OFDM}}_{n_a}}}^{T}]}^{T}.
\end{align*}
\end{small}         

\vspace{-2mm}
\hspace{-5mm}
The input-output relation for MIMO-OFDM can be written as
\begin{equation}
\mathbf{y}_{{\tiny \mbox{MIMO-OFDM}}} = \mathbf{H}_{{\tiny \mbox{MIMO-OFDM}}}\mathbf{x}_{{\tiny \mbox{MIMO-OFDM}}} + \mathbf{v}_{\tiny \mbox{MIMO-OFDM}}, 
\label{mimoofdmvecform}
\end{equation}
where 
$\mathbf{x}_{{\tiny \mbox{MIMO-OFDM}}},\mathbf{y}_{{\tiny \mbox{MIMO-OFDM}}}, \mathbf{v}_{\tiny \mbox{MIMO-OFDM}} \in \mathbb{C} ^{{n_a}NM \times 1}$
and
$\mathbf{H}_{{\tiny \mbox{MIMO-OFDM}}} \in \mathbb{C}^{{n_a}NM\times {n_a}NM}$.
  
\subsection{Performance results and discussions}
\label{sec4c}
In this subsection, we present the BER performance of MIMO-OTFS and compare 
it with that of MIMO-OFDM. Perfect channel knowledge is assumed at the 
receiver. Message passing algorithm is used for both MIMO-OTFS and MIMO-OFDM. 
A damping factor of 0.5 is used. The maximum number of iterations and the 
$\epsilon$ value used are 30 and 0.01, respectively. We use the channel 
model in (\ref{sparsechannelmimo}) and the number of taps $P$ is taken to 
be 5. The delay-Doppler profile considered in the simulation is shown in 
Table \ref{delay_Dopp_prof}. Other simulation parameters used are given in 
Table \ref{SimPar}.
\begin{table}
\begin{center}
\begin{tabular}{ |c|c|c|c|c|c| } 
\hline
Path index $(i)$ & $1$ &  $2$ &  $3$ &  $4$ &  $5$  \\ 
\hline
Delay ($\tau_i$), $\mu$s & $2.08$  & $4.164 $  & $6.246$  & $8.328$  & $10.41$\\
\hline
Doppler ($\nu_i$), Hz & $0$ & $470$  & $940$  & $1410$ & $1880$   \\
\hline
\end{tabular}
\vspace{2mm}
\caption{Delay-Doppler profile for the channel model with $P=5$.}
\vspace{-3mm}
\label{delay_Dopp_prof}
\end{center}
\end{table}
 
\begin{table}
\begin{center}
\begin{tabular}{|p{0.50\linewidth}|p{0.25\linewidth}|}
\hline
\textbf{Parameter} & \textbf{Value} \\
\hline
Carrier frequency (GHz) & 4 \\
\hline
Subcarrier spacing (kHz) &  15\\
\hline
Frame size $(M,N)$ & $(32,32)$ \\
\hline
Modulation scheme & BPSK \\
\hline
MIMO configuration & 1$\times$1, 2$\times$2, 3$\times$3 \\
\hline
Maximum speed (kmph) & 507.6\\
\hline
\end{tabular} 
\vspace{2mm}
\caption{System parameters.}
\label{SimPar}
\vspace{-9mm}
\end{center}
\end{table}

\begin{figure}
\hspace{4mm}
\includegraphics[width=9 cm, height= 6.0cm]{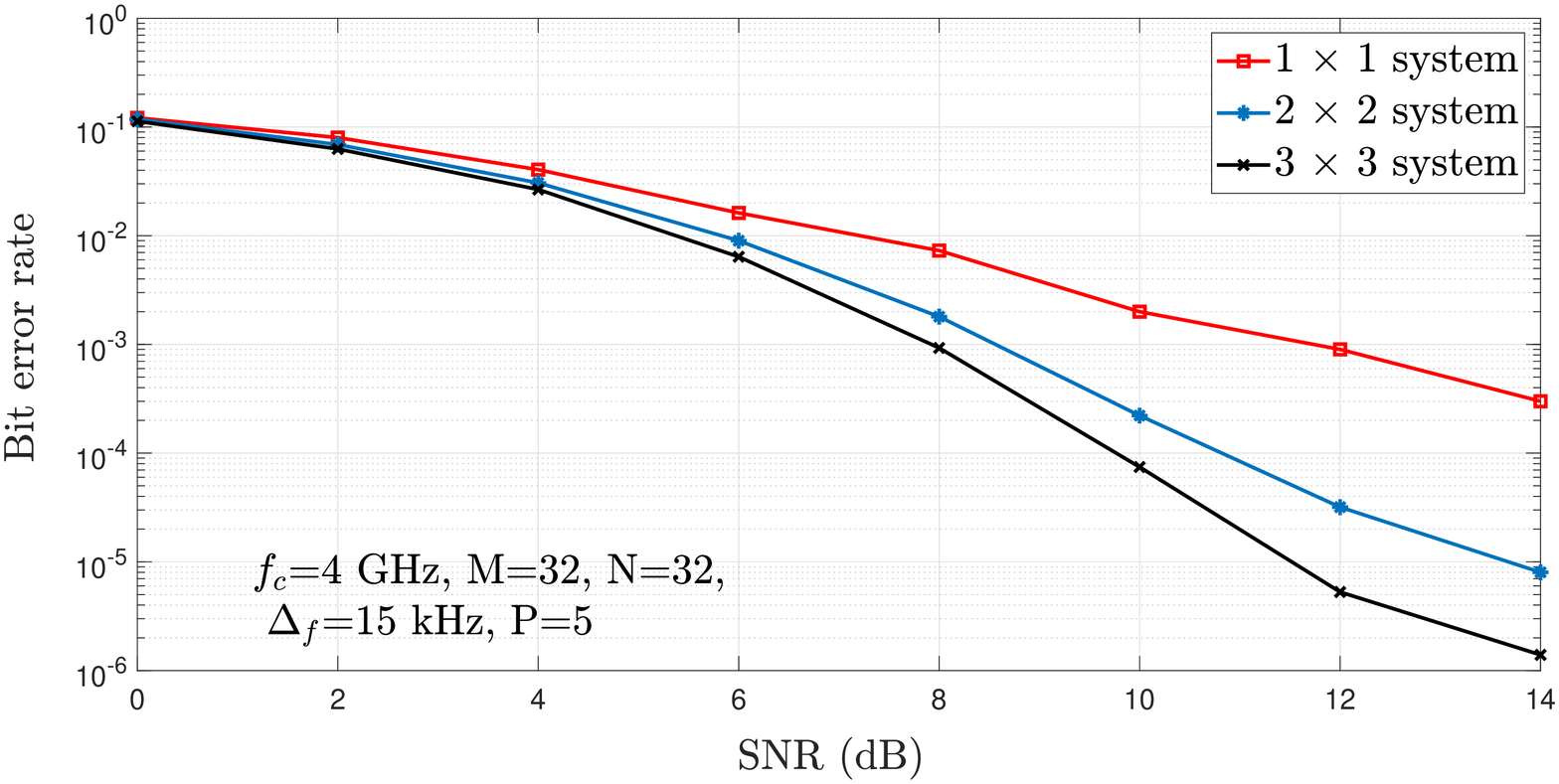}
\vspace{-6mm}
\caption{BER performance of MIMO-OTFS for SISO, and $2\times 2$ and 
$3\times 3$ MIMO systems.}
\label{BER1}
\vspace{-4mm}
\end{figure}

Figure \ref{BER1} shows the BER performance of MIMO-OTFS for SISO as well as
$2\times 2$ and $3\times 3$ MIMO configurations. The maximum considered 
speed of 507.6 kmph corresponds to 1880 Hz Doppler frequency at a carrier 
frequency of 4 GHz. Even at this high-Doppler value, MIMO-OTFS is found to 
achieve very good BER performance. We observe that, a BER of $10^{-5}$ is 
achieved at an SNR of about 14 dB for the 2$\times$2 system, while the SNR 
required to achieve the same BER reduces by about 2 dB for the 3$\times 3$ 
system. Thus, with the proposed detection algorithm, MIMO-OTFS brings in 
the advantages of linear increase in spectral efficiency with number of 
transmit antennas and the robustness of OTFS modulation in high-Doppler 
scenarios.

\begin{figure}
\hspace{4mm}
\includegraphics[width=9 cm, height= 6.0cm]{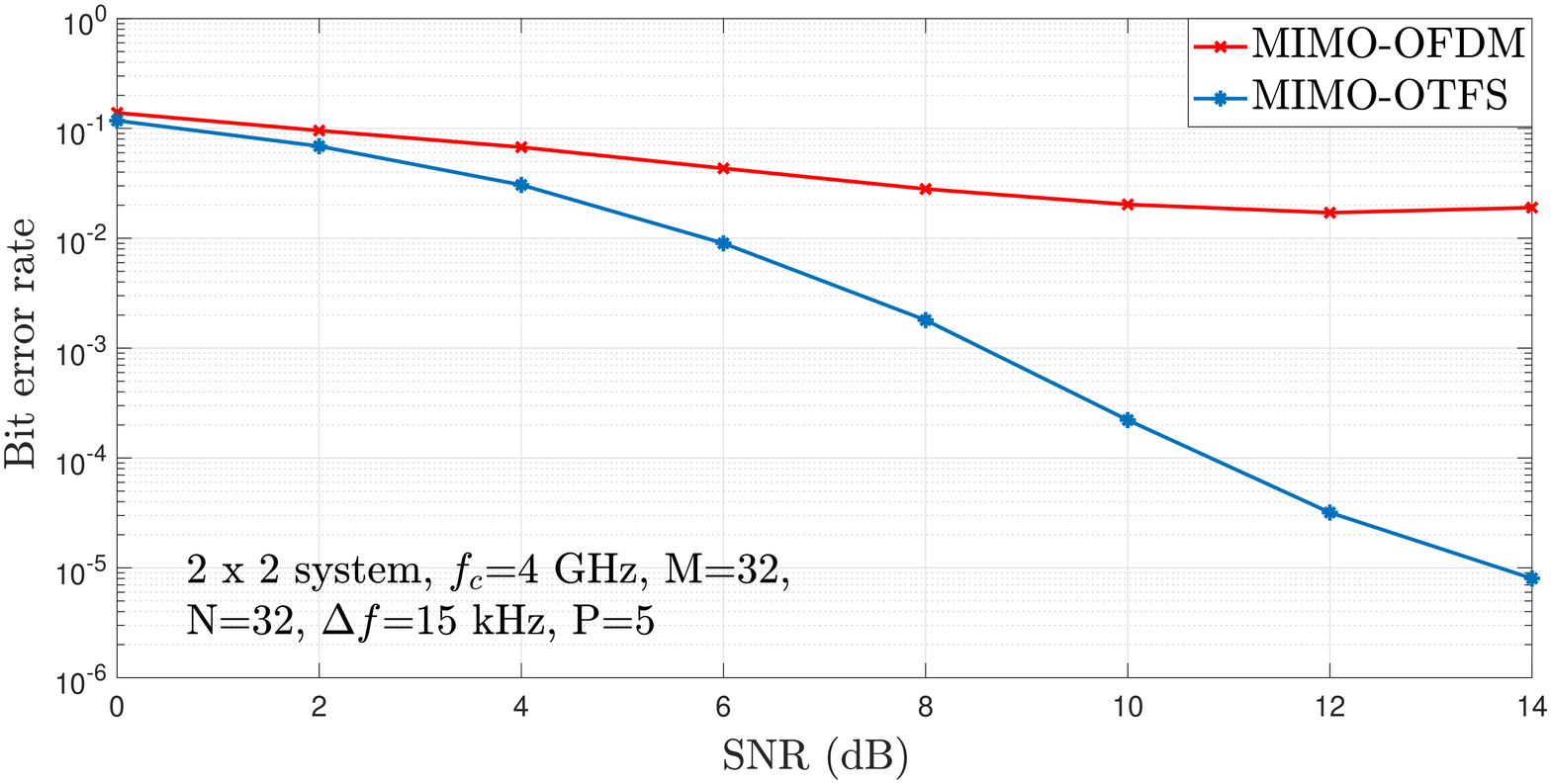}
\vspace{-6mm}
\caption{BER performance comparison between MIMO-OTFS and MIMO-OFDM in
a $2\times 2$ MIMO system.}
\label{BER2}
\vspace{-3mm}
\end{figure}

\begin{figure*}[t]
\centering
\includegraphics[scale=0.32]{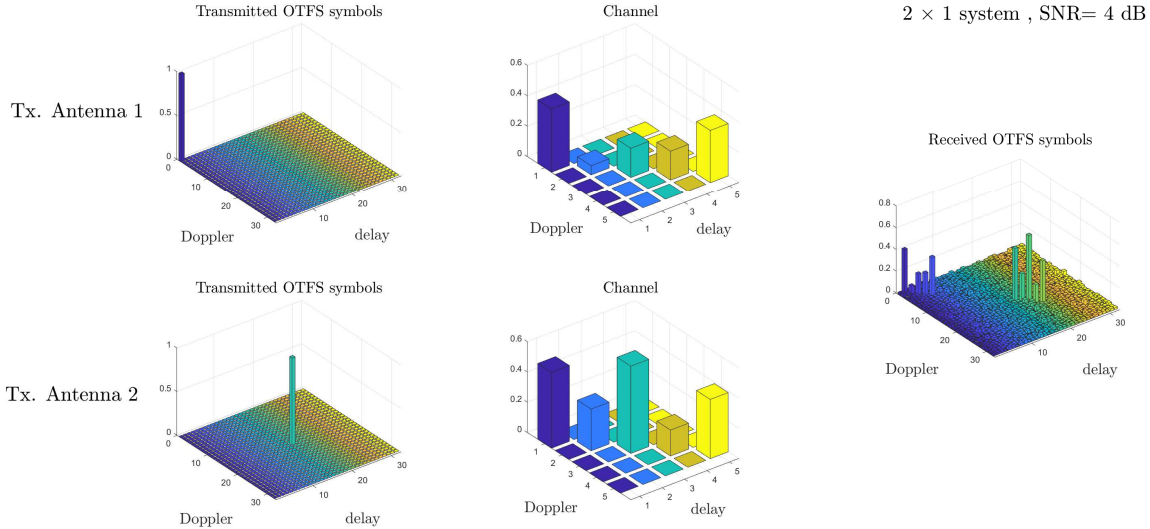}
\vspace{-6mm}
\caption{Illustration of pilots and channel response in delay-Doppler 
domain in a 2$\times$1 MIMO-OTFS system.}
\label{CSE_MIMO}
\vspace{-4mm}
\end{figure*}

Figure \ref{BER2} shows the BER performance comparison between MIMO-OTFS 
and MIMO-OFDM in a $2\times 2$ MIMO system. The maximum Doppler spread in 
the considered system is high (1880 Hz) which causes severe ICI in the TF 
domain. Because of the severe ICI, the performance of MIMO-OFDM is found 
to break down and floor at a BER value of about $2 \times 10^{-2}$. However, 
MIMO-OTFS is able to achieve a BER of $10^{-5}$ at an SNR value of about 
14 dB. This is because OTFS uses the delay-Doppler domain for signaling 
instead of TF domain. Thus, the BER plots clearly illustrate the robust 
performance of MIMO-OTFS and its superiority over MIMO-OFDM under rapidly 
varying channel conditions.    

\section{Channel Estimation for MIMO-OTFS}
\label{sec5}
In this section, we relax the assumption of perfect channel knowledge and 
present a channel estimation scheme in the delay-Doppler domain. The scheme 
uses impulses in the delay-Doppler domain as pilots. Figure \ref{CSE_MIMO} 
gives an illustration of the pilots, channel response, and received signal 
in a $2\times 1$ MIMO system with the delay-Doppler profile and system 
parameters given in Tables \ref{delay_Dopp_prof} and \ref{SimPar}. Each 
transmit and receive antenna pair sees a different channel having a finite 
support in the delay-Doppler domain. The support is determined by the delay 
and Doppler spread of the channel \cite{otfs1}. This fact can be used to 
estimate the channel for all the transmit-receive antenna pairs 
simultaneously using a single MIMO-OTFS frame as described below. 

The OTFS input-output relation  for $p$th transmit antenna and $q$th 
receive antenna pair can be written using (\ref{otfsinpoutp}) as

\vspace{-2mm}
\begin{small}
\begin{equation}
{\hat{x}}_{q}[k,l]= \sum_{m=0}^{M-1} \sum_{n=0}^{N-1} x_{p}[n,m] {1 \over MN} {h_{w_{qp}}} \left( {k-n \over NT}, {l-m \over M \Delta f} \right)+v_q[k,l].
\label{otfsinpoutpmimo}
\end{equation}
\end{small}

\vspace{-3mm}
\hspace{-4mm}
If we transmit 
\begin{align}
x_{p}[n,m]&=1 \  \text{if}\  (n,m)=(n_{p},m_{p}) \nonumber\\
&=0 \ \forall \ (n,m) \neq (n_{p},m_{p}),
\end{align}  
as pilot from the $p$th antenna, the received signal at the $q$th antenna 
will be
\begin{equation}
\label{estimate}
{\hat{x}}_{q}[k,l]={1 \over MN} {h_{w_{qp}}} \left( {k-n_p \over NT}, {l-m_p \over M \Delta f} \right)+v_q[k,l].
\end{equation}
We can estimate 
${1 \over MN} {h_{w_{qp}}} \left( {k \over NT}, {l \over M \Delta f} \right)$ 
from (\ref{estimate}), since, being the pilots, $n_p$ and $m_p$ are known at 
the receiver a priori. From this, we can get the equivalent channel matrix 
$\hat{\mathbf{H}}_{qp}$ using the vectorized formulation of Sec. \ref{sec2c}.  
From (\ref{estimate}) we also see that, due to the 2D-convolution input-output 
relation, the impulse at $(n,m)=(n_{p},m_{p})$ is spread by the channel only 
to the extent of the support of the channel in the delay-Doppler domain. Thus, 
if we send the pilot impulses from the transmit antennas with sufficient 
spacing in the delay-Doppler domain, they will be received without overlap. 
Hence, we can estimate the channel responses corresponding to all the 
transmit-receive antenna pairs simultaneously and get the estimate of the 
equivalent MIMO-OTFS channel matrix $\hat{\mathbf{H}}_{{\tiny \mbox{MIMO}}}$ 
using a single MIMO-OTFS frame. This is illustrated in Fig. \ref{CSE_MIMO} 
for a $2\times 1$ MIMO-OTFS system with frame size $(M,N)=(32,32)$ at an SNR 
value of 4 dB. The first antenna transmits the pilot impulse at 
$(n_1,m_1)=(0,0)$ and the second antenna transmits the pilot impulse at 
$(n_2,m_2)=(16,16)$ in the delay-Doppler domain. We observe that the impulse 
response ${h_{w_{11}}}\left( {k-n_1 \over NT},{l-m_1\over M \Delta f} \right)$ 
and ${h_{w_{12}}} \left({k-n_2 \over NT}, {l-m_2 \over M \Delta f} \right)$  
are non-overlapping at the receiver. Thus, they can be estimated 
simultaneously using a single pilot MIMO-OTFS frame.
 
\subsection{Performance results and discussions}
\label{sec5c}
In this subsection, we present the BER performance of the MIMO-OTFS system 
using the estimated channel. We use the MIMO-OTFS channel estimation scheme 
described above, for estimating the equivalent channel matrix 
$\hat{\mathbf{H}}_{{\tiny \mbox{MIMO}}}$ and use the message passing 
algorithm for detection. The delay-Doppler profile and the simulation 
parameters are as given in Table \ref{delay_Dopp_prof} and Table 
\ref{SimPar}, respectively.

\begin{figure}
\hspace{4mm}
\includegraphics[width=9cm, height= 6.0cm]{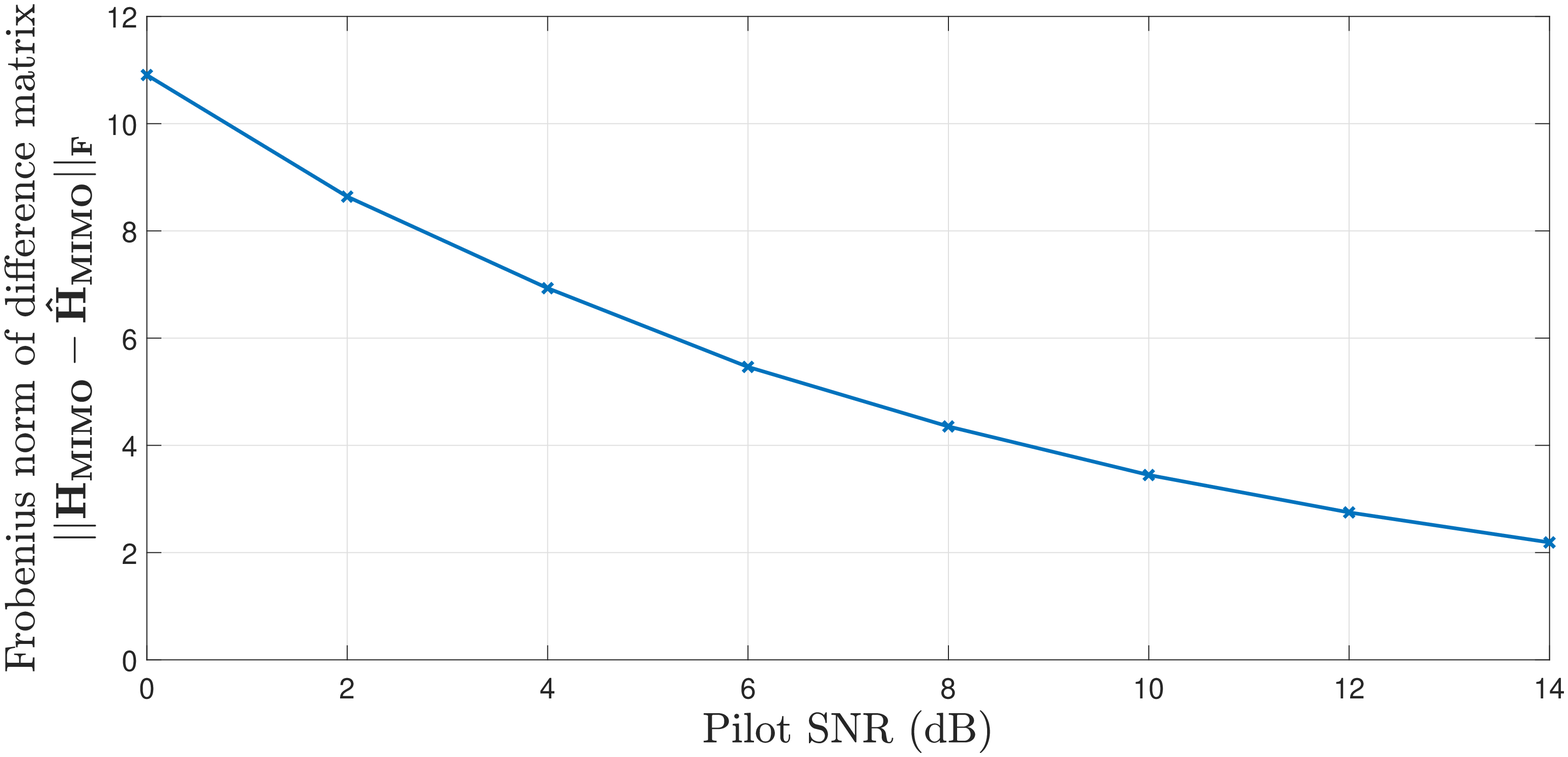}
\vspace{-7mm}
\caption{Frobenius norm of the difference between the equivalent channel 
matrix (${\mathbf{H}}_{{\tiny \mbox{MIMO}}}$)  and the estimated equivalent 
channel matrix ($\hat{\mathbf{H}}_{{\tiny \mbox{MIMO}}}$) as a function of 
pilot SNR in a 2$\times$2 MIMO-OTFS system.}
\label{est_err_p_snr}
\vspace{-2mm}
\end{figure}

\begin{figure}
\hspace{4mm}
\includegraphics[width=9cm, height= 6.0cm]{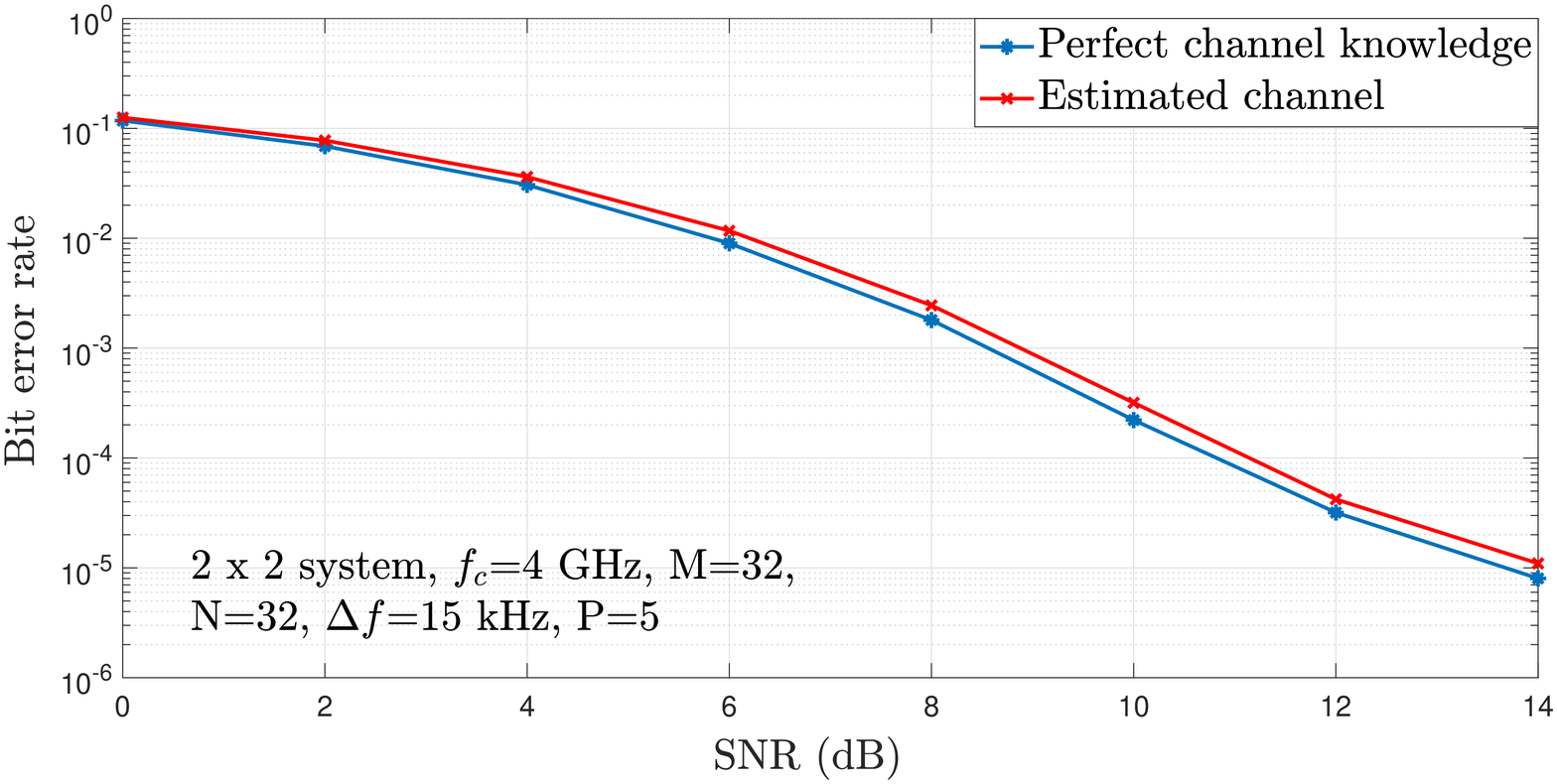}
\vspace{-5mm}
\caption{BER performance of MIMO-OTFS system using the estimated channel 
in a 2$\times$2 MIMO-OTFS system.}
\label{BER3}
\vspace{-4mm}
\end{figure}

In Fig. \ref{est_err_p_snr}, we plot the Frobenius norm of the difference 
between the equivalent channel matrix (${\mathbf{H}}_{{\tiny \mbox{MIMO}}}$) 
and the estimated equivalent channel matrix 
($\hat{\mathbf{H}}_{{\tiny \mbox{MIMO}}}$) (a measure of estimation error) 
as a function of pilot SNR for a $2\times 2$ MIMO-OTFS system with system 
parameters as in Tables \ref{delay_Dopp_prof} and \ref{SimPar}. We observe 
that, as expected, the Frobenius norm of the difference matrix decreases 
with pilot SNR. Figure \ref{BER3} shows the corresponding BER performance 
using the proposed channel estimation scheme for the $2\times 2$ MIMO-OTFS 
system. It is observed that the BER performance achieved with the estimated 
channel is quite close to the performance with perfect channel knowledge. 
For example, a BER of $2\times 10^{-5}$ is achieved at SNR values of about 
12.5 dB and 13 dB with perfect channel knowledge and estimated channel 
knowledge, respectively. At the considered maximum Doppler frequency of 
1880 Hz, channel estimation in the time-frequency domain leads to inaccurate 
estimation because of the rapid variations of the channel in time. On the 
other hand, the sparse channel representation in the delay-Doppler domain 
is time-invariant over a larger observation time. This, along with the OTFS 
channel-symbol coupling (2D periodic convolution) in the delay-Doppler 
domain, enables the proposed channel estimation for MIMO-OTFS to be simple 
and efficient. 

\section{Conclusions}
\label{sec6}
We investigated signal detection and channel estimation aspects of MIMO-OTFS 
under high-Doppler channel conditions. We developed a vectorized formulation 
of the input-output relationship for MIMO-OTFS which enables MIMO-OTFS signal 
detection. We presented a low complexity iterative algorithm for MIMO-OTFS 
detection based on message passing. The algorithm was shown to achieve very 
good BER performance even at high Doppler frequencies (e.g., 1880 Hz) in a 
$2\times 2$ MIMO system where MIMO-OFDM was shown to floor in its BER 
performance. We also presented a channel estimation scheme in the 
delay-Doppler domain, where delay-Doppler impulses are used as pilots. The 
proposed channel estimation scheme was shown to be efficient and the BER 
degradation was small as compared to the performance with perfect channel 
knowledge. The sparse nature of the channel in the delay-Doppler domain 
which is time-invariant over a larger observation time enabled the proposed 
estimation scheme to be simple and efficient.

\end{document}